# Formation of Epitaxial MnBi Layers on (Ga,Mn)As

J. Adell[1], M. Adell[1], I. Ulfat[1], L. Ilver[1], J. Sadowski[2,3]
and J. Kanski[1]

[1] *Department of Applied Physics, Chalmers University of Technology, SE-412 96 Göteborg, Sweden*
[2] *MAX-lab, Lund University, SE-221 00 Lund, Sweden*
[3] *Institute of Physics, Polish Academy of Sciences, al. Lotnikow 32/46, 02-668 Warszawa, Poland*

(Date: 5 March 2009)

The initial growth of MnBi on MnAs terminated (GaMn)As is studied by means of synchrotron based photoelectron spectroscopy. From analysis of surface core level shifts we conclude that a continued epitaxial MnBi layer is formed, in which the MnAs/MnBi interface occurs between As and Bi atomic planes. The well defined 1x2 surface reconstruction of the MnAs surface in preserved for up to 2 ML of MnBi before clear surface degradation occurs.

## I. INTRODUCTION

The interest in diluted magnetic semiconductors remains high and (Ga,Mn)As continues to play the important role of a model system for investigation of electronic and magnetic properties.[1,2] For some time it was believed that the Curie temperature of this system would be limited to around 110K for fundamental (but unknown) reasons. Eventually the negative role of point defects, especially Mn atoms in interstitial sites, was recognized, and it was found that these particular defects could be removed by post-growth annealing. Thus, several groups have been able to produce (Ga,Mn)As layers with ferromagnetic transition temperatures in the range 160-180 K.[3,4] The limiting mechanisms are still not fully understood but can be suspected to be of experimental rather than of fundamental nature, in which case Curie temperatures above RT may eventually be reached. According to the RKKY model the Curie temperature is proportional to the density of charge carriers and to the density of magnetic ions, so an obvious way to reach higher Curie temperatures is by increasing Mn concentration while keeping the defect concentration at a minimum. Unfortunately, to raise the Mn content in (Ga,Mn)As the growth temperature must be lowered, and this in turn promotes generation of point defects. An alternative, still unexplored approach is to integrate ultra thin epitaxial layers of ferromagnetic materials, e.g. MnAs, MnSb, or MnBi. Calculations[5] predict that these materials should exhibit half-metallic ferromagnetism in the zincblende (ZB) structure and that MnBi is expected to remain a ferromagnetic half-metal in the tetragonally strained configuration expected for an epitaxial layer on GaAs.[6]

As has been demonstrated earlier,[7] post-growth annealing of (Ga,Mn)As under As capping is very efficient to eliminate interstitial Mn from the as-grown layer and thereby increase the ferromagnetic transition temperature. As a by-product of such treatment the annealed (Ga,Mn)As becomes terminated by 1-2 monolayers of epitaxial MnAs. It will be shown here that starting from such a surface it is possible to generate a well-ordered continuous epitaxial MnBi layer.

## II. EXPERIMENT

The experiments were performed at beamline 41 at the Swedish national synchrotron radiation laboratory MAX-lab, where a Molecular Beam Epitaxy (MBE) system is connected to the photoelectron spectrometer for on-line sample preparation. The samples were grown on 1x1 cm$^2$ pieces of n-type epiready GaAs(100) wafers, which were glued by In to Mo sample holders. The substrate temperature was measured with an IR pyrometer and the Ga and Mn beam fluxes were calibrated by means of RHEED oscillations.[8] After conventional oxide desorption and growth of a GaAs buffer at 580 ºC, the substrate temperature was reduced to 230ºC for growth of a 500 Å thick Ga$_{0.95}$Mn$_{0.05}$As layer. The surface of this as-grown layer showed a weak but clear 1x2 reconstruction. After annealing at 230ºC under As capping the half order diffraction streaks were significantly enhanced. Bi was deposited on this surface at room temperature (RT). The Bi coverage was estimated to reach one monolayer when the RHEED pattern was vanishing. Subsequent core level analysis indicated that the dosing was somewhat in excess of one monolayer. A clear 1x2 RHEED pattern was restored after a new annealing treatment at 195 ºC for 90 min. On this surface additional monolayers of Mn and Bi were deposited sequentially, Mn at 200°C and Bi at RT followed by 90 min annealing at 195 °C. In this way altogether three Bi and two Mn layers were deposited.

Angle resolved photoemission from the shallow core levels of As and Bi was used to follow the development of adsorbed layers. The photoemission spectra were recorded in the plane of light incidence, with the p-polarized light incident at 45° relative surface normal. It is important to stress that the samples were transferred between the growth and analysis chambers under UHV conditions. The direct connection of the two systems allows us to follow the sample development after sequential growth steps. Most significantly, it allows us to study well defined as-grown samples without any need to restore the surface. As the material is metastable and undergoes phase separation around 300 °C, its surface cannot be restored to a well ordered state once it has been exposed to air.

### III. RESULTS AND DISCUSSION

The overlayers were deposited in the MBE system and the surfaces were monitored during growth by means of RHEED. Although some degradation could be noticed after the third Mn and Bi deposition cycle, up to that point a streaky and low background 1x2 pattern was observed, typical for a flat and well-ordered surface. After transfer to the spectrometer chamber the surface order was confirmed by LEED.

A continuous layer growth could also be concluded from the vacuum level cut-off in the photoemission data. In Fig. 1 we show the development of the low-energy range of the spectra, which are displayed on a relative energy scale with reference to the initial MnAs-terminated situation (a). After the first Bi deposition the cut-off is lowered by 0.50 eV (spectrum b). Apart from the shift, it is noted that the spectrum does not show any secondary structure at the position of the initial cut-off. We can, therefore, conclude that the surface is essentially free from uncovered patches and that the streaky RHEED pattern of this surface does indeed represent the Bi covered surface and not uncovered regions of the initial MnAs-terminated (Ga,Mn)As surface. After subsequent deposition of a monolayer Mn the vacuum cut-off shifts further by 0.25 eV (spectrum c), and another layer of Bi gives an additional shift of 0.07 eV (spectrum d). In the following deposition cycle of Mn and Bi monolayers the cut-off returns to positions c and d, respectively.

We note that spectrum d) does not show the typical monotonous decay towards increasing energy, but has a broad structure in the range of the cut-off of the Bi-terminated MnAs surface. We interpret this structure as a secondary cut-off due to surface regions covered with excess Bi. By subtracting spectrum b from spectrum d (after adequate shift) we find that the difference, shown by small symbols in Fig. 1, has a low-energy limit rather close to spectrum b). The two systems are apparently terminated in similar ways, with a double layer of column V elements on top of a Mn layer. Thus, the data indicate that Bi adsorbed on the MnAs-terminated surface does not break up the Mn-As bonds but forms an ordered layer on top of the As atoms.

The work function changes reflect modifications of the surface dipole, i.e. charge redistributions between the outermost surface planes. Using the most common electronegativity scale of Pauling, according to which the electronegativities of As, Bi, and Mn are respectively 2.18, 2.02, and 1.55, one should thus expect a slightly reduced work function after the first Bi adsorption. Therefore it is somewhat surprising to see that the work function of the Bi-terminated surface is reduced by as much as 0.5 eV, and that the following reduction after Mn adsorption is smaller (0.25 eV) despite the clearly lower electronegativity of Mn. One obvious reason for this result is that the surface dipole depends not only on the last two atomic planes. In the present case the As atoms in the MnAs layer are polarized before the Bi adsorption due to interaction with Mn, and therefore the atomic electronegativity of As is not an appropriate parameter. The relatively small and reversible changes found after the last Mn and Bi depositions suggest that the effective electronegativity difference between Bi and Mn is relatively small. On alternative electronegativity scales that take into account the atomic size (Allred Rochow and Sanderson scales) the electronegativities of Bi (1.67 and 2.34) are indeed much closer to those of Mn (1.60 and 2.20) than to As (2.20 and 2.82), though in all cases Bi is somewhat more negative. Nevertheless, in all cases Mn is less electronegative than Bi, which should give surface dipoles of opposite signs than those found here if only the two last layers were active. A more detailed quantitative analysis of the charge distribution would be needed to describe the work function changes observed here. Such analysis is beyond the scope of this work.

The formation of the MnBi overlayer can also be followed qualitatively by the development of photoemission survey spectra, see Fig. 2. After the first Bi deposition and annealing at 195 °C (spectrum b) the As/Ga intensity ratio is reduced due to the removal of loosely adsorbed surface

As. With a Mn layer adsorption and a second Bi deposition and annealing (spectrum c) both Ga and As peaks are reduced relative to the Bi and Mn emission. The apparently stronger attenuation of the Ga peak is explained by the shorter elastic mean free path in the kinetic energy range 50-60 eV than around 35 eV. Thus, the As and Ga signals decrease systematically as we add monolayers of Bi and Mn, just as expected from layer-by-layer coverage.

To get a more detailed picture of the overlayer character we analyze the As3d and Bi5d core level spectra. We start with As, which is the more complicated case due to a number of strongly overlapping components. The spectral decomposition presented in Fig. 3 is based on requirements of consistency between spectra obtained at different emission angles and under different surface conditions. In addition, we have also made use of a recent analysis of spectra from As-annealed (Ga,Mn)As, in which we were eventually able to identify the component representing the MnAs surface layer.[9] The fitting parameters are summarized in Table 1. We find that a consistent decomposition of the spectrum obtained from a sample annealed under As can be achieved with five components, which we assign as atoms in the (Ga,Mn)As bulk (B), adsorbed As (A), two surface components (S1, S2) and a fifth component (X) which derives from the reacted MnAs layer.[9] The spectrum strongly resembles that from as-grown (Ga,Mn)As,[10] though a direct comparison shows immediately that the spectra differ with respect to their total widths. By comparing the angular and thermal dependences we can find the corresponding components in the two spectra. In particular we find that component X (see Fig. 3) is more stable with respect to annealing than any of the components found on the as-grown surface (except for the bulk component). The increased width of the spectrum from the MnAs-terminated is caused by a clearly larger separation between components B and A (1.0 eV vs. 0.87 eV), indicating a larger binding energy of the adsorbed species on the MnAs-terminated surface. The energy separation B-S1 is very similar on the as-grown and MnAs-terminated surfaces (around 0.30 eV), while B-S2 is again significantly larger for the MnAs terminated surface (0.76 eV vs 0. 45 eV).

With adsorption of the first monolayer Bi some changes are observed in the As 3d spectrum. Apart from an overall intensity reduction, seen in Fig. 2, the relative intensities of the different components are affected. The most apparent change is the reduction of component A, i.e. that reflecting adsorbed As. However, as it is not completely removed, we conclude that some As is floating on top of the Bi-covered surface. This is most likely unintentionally adsorbed As in the growth chamber atmosphere. Furthermore, we note that components S1 and S2 also remain, confirming the presence of As on the surface. The intensity of S1 is actually increased relative B, which is of course mainly a consequence of attenuation of B by the Bi overlayer. The most significant observation is that the intensity of X is practically unchanged. This component clearly reflects atoms in a stable bonding state, supporting our earlier assignment as As-atoms in the MnAs layer. We conclude that the adsorbed Bi atoms bind to As in the MnAs layer, as already inferred from the work function data. The adsorption of Bi results in a slight shift of the X component towards higher binding energy as indicated in Fig. 3.

After deposition of 2 ML MnBi an intensity reduction is observed in the As 3d emission (c.f. Fig. 2). The main change in the component-resolved analysis spectrum is a complete removal of component A. S1 and S2 are still present, though S2 with even lower intensity, showing that some As is still present on the surface. The MnAs-related component X remains with approximately unchanged relative intensity - the small reduction (about 10%) relative the preceding spectrum may be due to diffraction effects or even an artifact in the fitting procedure.

We next turn to the apparently less complex Bi 5d emission. Spectra obtained at different stages of treatment are shown in Fig. 4. The spectra are normalized to approximately the same amplitude and the energy is given relative the fermi energy, which is assumed to be unaffected by the surface treatments. Directly after RT adsorption (spectrum "a") two components are found, separated by

0.36 eV. The presence of two components is tentatively ascribed to Bi adsorbed on the As-terminated surface in different configurations. Bi adsorbed in the subsequent steps on Mn-terminated surfaces was characterized by a single Bi component (see below). Upon annealing the intensity is clearly redistributed. Again two components are found, separated in energy by 0.49 eV. This is clearly larger than before annealing, and shows that some reaction has taken place. The smaller component (A) coincides with the main one in the previous spectrum, and is thus ascribed to adsorbed, unreacted Bi. Its presence shows that the Bi dosing was somewhat higher than intended. The main component is consequently associated with reacted Bi, i.e. atoms bonding to As in the surface layer. In the subsequent step the sample was covered with 1.0 ML Mn (the Mn flux was calibrated by means of RHEED oscillations during growth of the underlying (GaMn)As). This treatment leads to some intensity reduction, but leaves otherwise the main component unaffected. The spectrum tails somewhat on the high energy side, which can be accounted for by introducing a third Bi component. As this component does not appear in any later spectra, we are not able to identify its origin. The relative increase of component A suggests that the excess Bi is floating on top of the Mn layer. This is supported by the development with further Bi adsorption. We see in spectrum d) that the intensity of component A is enhanced after RT deposition of another Bi layer, and under the more surface sensitive conditions at 60° emission angle (not shown) this component is clearly dominating.

Annealing of this surface results in spectrum e). As expected, the adsorbed component is reduced, but in comparison with spectrum b), which was obtained after equivalent annealing, spectrum e) contains an additional component on the high energy side. Considering the preparation history Bi should occur in at least two inequivalent sites: i) atoms from the first deposition should be mainly located between the last substrate As layer and the Mn layer deposited in step c), i.e. interface sites, and ii) Bi atoms from the second adsorption cycle, now forming the surface layer reacted with Mn. It is natural to expect the surface sites to dominate, so we ascribe the largest of the two peaks to these sites. We also note that the smaller of the two components is slightly shifted (by about 80 meV) relative component A in the previous spectrum. Another observation is that the spectrum recorded at 60° emission angle shows only a slight change in the relative intensity of the two components. These observations, along with the subsequent development with a new Mn adsorption, strongly suggest that the smaller of the two components in spectrum e) actually contains contributions from the interface atoms as well as atoms in an adsorbed state due to excess Bi. Such an arrangement would tend to compensate the expected angular dependence.

Following the second Mn adsorption (spectrum f), we see that the total width is further increased somewhat. The decomposition reveals the appearance of a third component, which is not expected considering that no new Bi has been supplied. We can explain the new component by assuming, as above, that some excess Bi above one monolayer was added in the latest deposition steps. By adding Mn, we thus not only transform the previous surface Bi into sites sandwiched between two Mn planes (B), but have in addition a new partial surface layer of reacted Bi (since the Mn adsorption was done at 200°C) on top of the last Mn layer. The identification of the new surface component S is based on the fact that this component is enhanced relative the others at off-normal emission. We also note that the other two components (B and I) appear with similar intensities also in off-normal emission. As above, we ascribe this to the coinciding emission from interface Bi and unreacted Bi floating on the surface.

The third RT adsorption of Bi resulted again in increased emission on the high-energy side of the spectrum, at the location of unreacted adsorbed Bi. The S component is reduced clearly more than B, which we believe is due to partial incorporation of the surface Bi into the adsorbed unreacted layer. However, after new annealing (spectrum h) the surface component is significantly enhanced, and the spectrum contains three components as described above. The angular dependence (Fig. 5) supports the assignment of the surface component, but also shows that component A/I (emission from adsorbed and interface Bi) is relatively strong in large emission angles. It is clear that the conversion from adsorbed to reacted surface Bi is incomplete under the present conditions.

## IV. SUMMARY

We have demonstrated that MnAs terminated (GaMn)As layers can be overgrown with at least two monolayers MnBi using alternating Bi and Mn deposition and annealing. The MnBi layer maintains the same 1x2 surface reconstruction as the GaMnAs surface and is apparently coherent with the underlying substrate. It seems, however, that incorporation of the third adsorbed Bi layer into the developing MnBi overlayer is hampered, probably as a result of lattice mismatch.

## V. ACKNOWLEDGEMENTS

The present work is part of a project supported by the Swedish Research Council (VR).

**Table 1:** Fitting parameters for As 3d and Bi $5d_{5/2}$ spectra.

|  | As 3d | Bi $5d_{5/2}$ |
|---|---|---|
| $\Delta E_{s\text{-}o}$ | 0.695 eV | - |
| Branching ratio | 1.69 | - |
| $\Delta E_{Gauss}$ | 0.46 eV | 0.63 eV |
| $\Delta E_{Lorentz}$ | 0.17 eV | 0.27 eV |

FIGURE CAPTIONS

Fig. 1  Development of the low-energy cut-off range with Bi and Mn adsorption. a) MnAs terminated GaMnAs, b) after deposition of 1ML Bi and annealing, c) after adsorption of 1 ML Mn, and d) after a second adsorption of Bi and annealing.

Fig. 2: Survey spectra obtained in normal emission excited with 81 eV photons. a) the starting MnAs-terminated surface b) after 1ML Bi deposition and annealing, and c) after the second deposition and annealing (corresponding to spectrum d in Fig. 1).

Fig. 3: Normal emission As3d spectra. a) the initial MnAs-terminated surface, b) after Bi deposition and annealing, and c) after additional Mn and Bi deposition and annealing.

Fig. 4: Normal emission $Bi5d_{5/2}$ spectra during growth progress. First row represents Bi deposited at room temperature, second row after annealing and final row after Mn deposition at 200°C. Each column represents a Bi and Mn deposition cycle.

Fig. 5: $Bi5d_{5/2}$ in normal and 60° emission angle after third Bi deposition and annealing. The angular dependence clearly supports our assignment of the surface component.

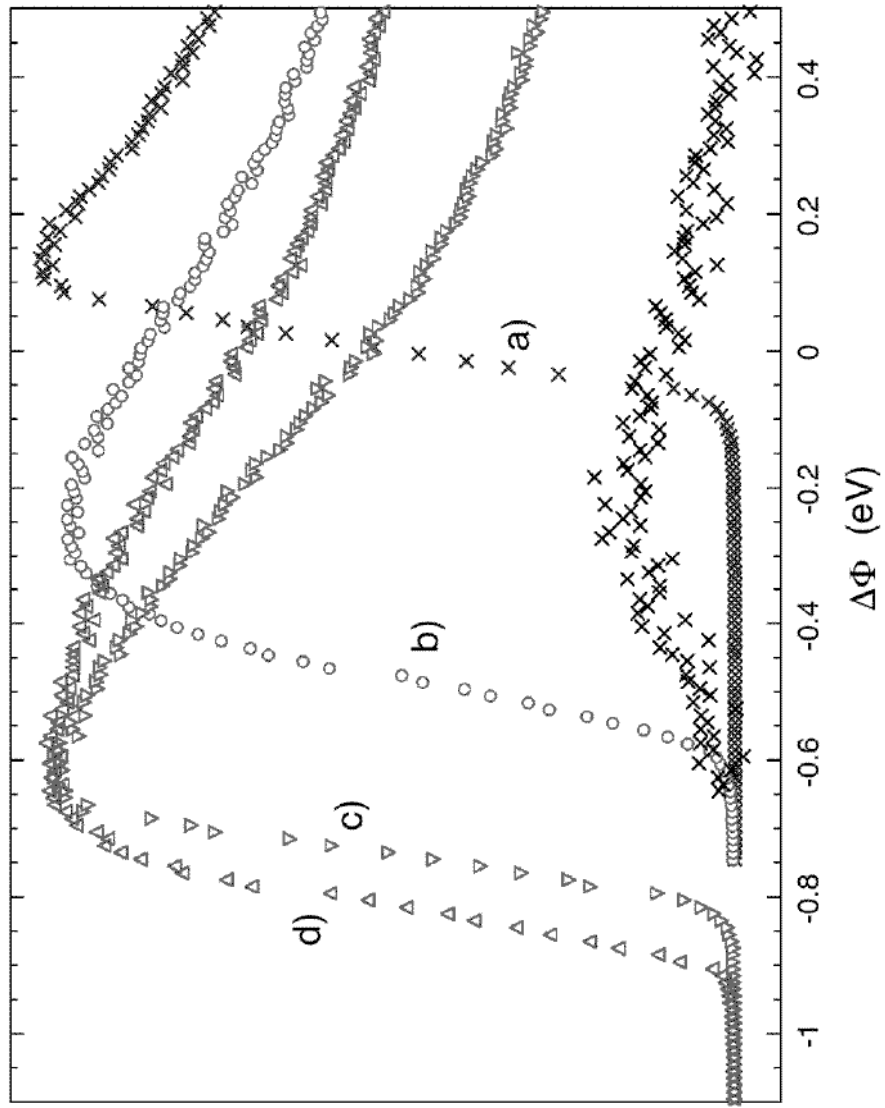

Figure 1

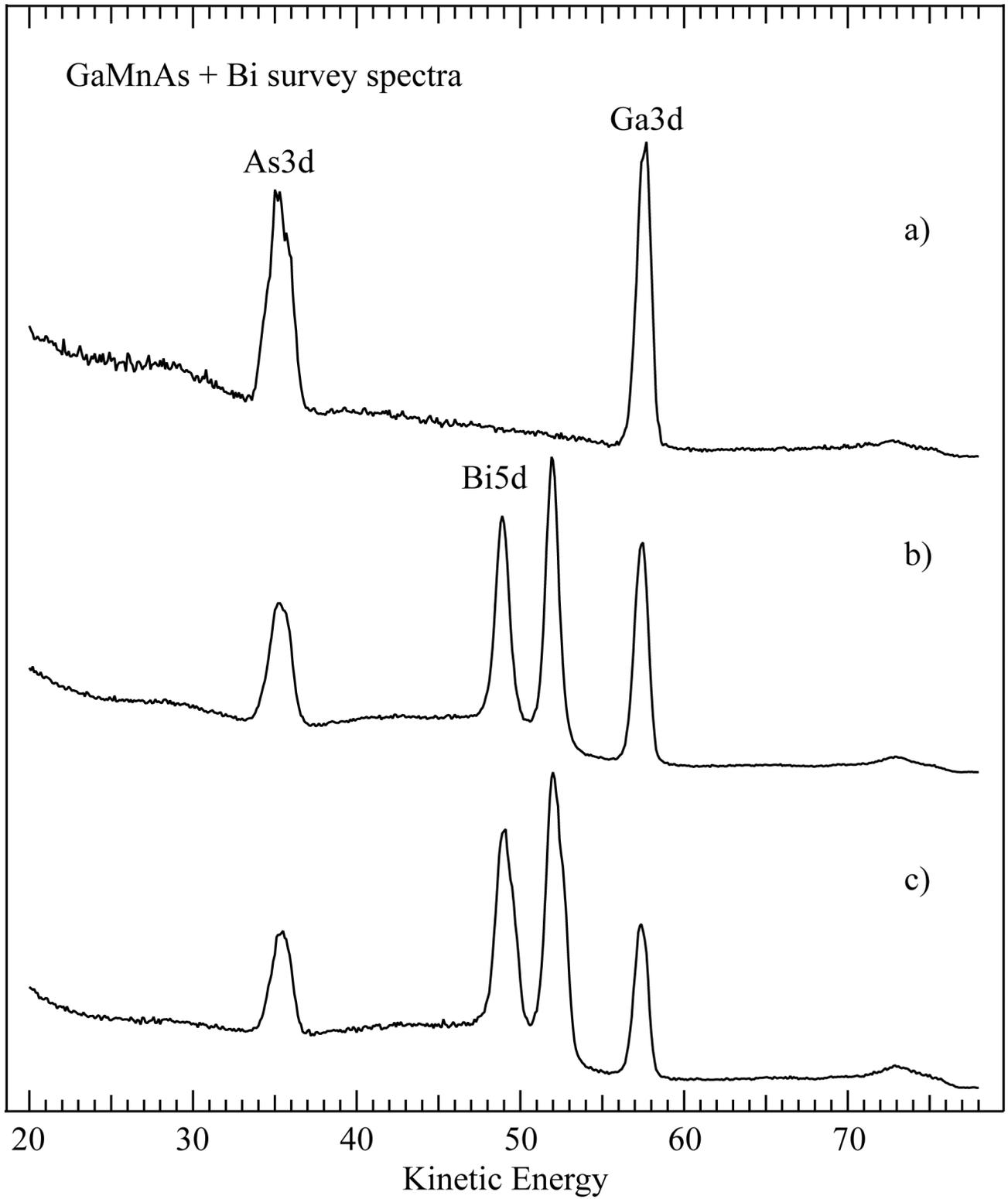

**Figure 2**

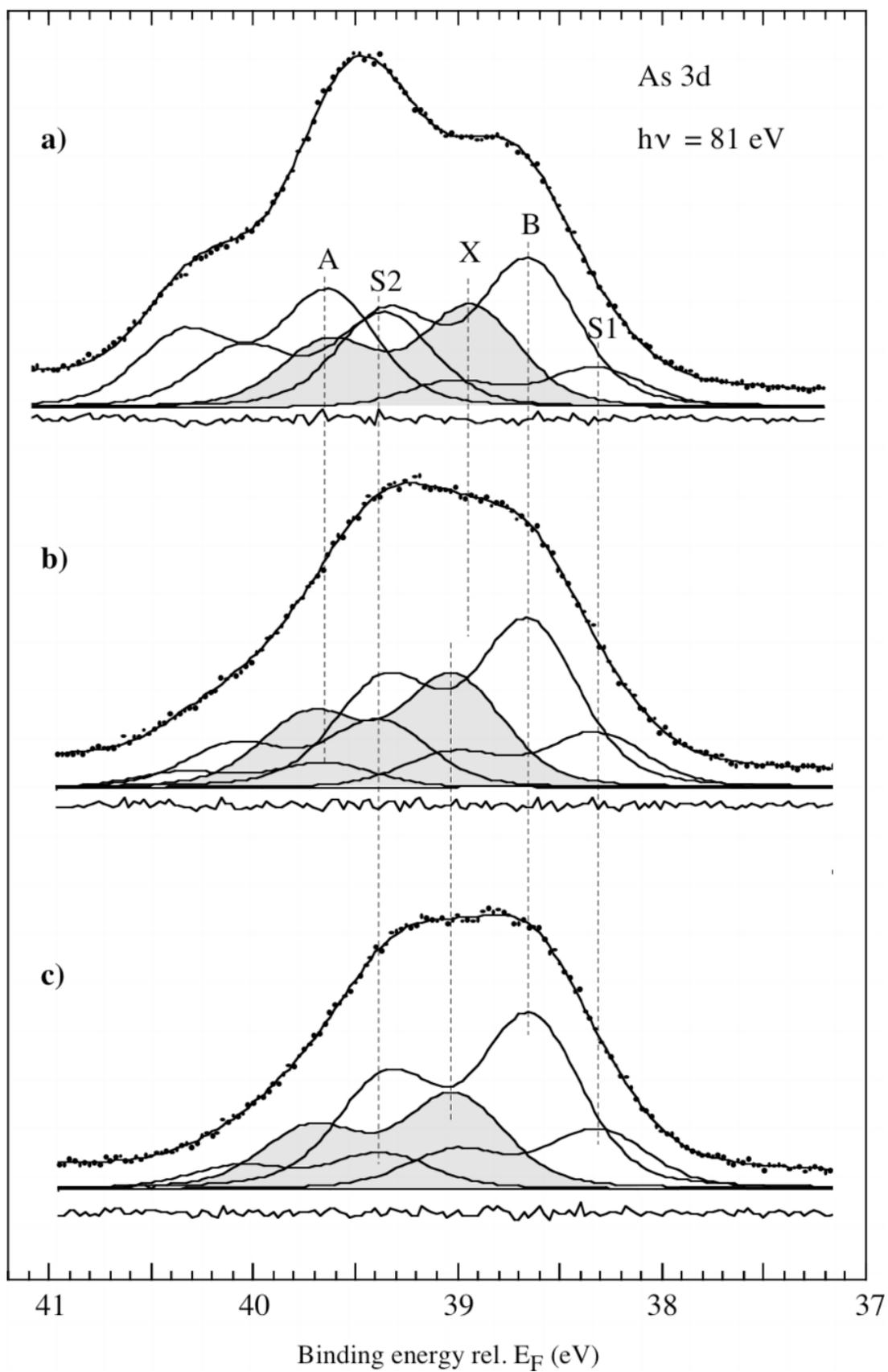

**Figure 3**

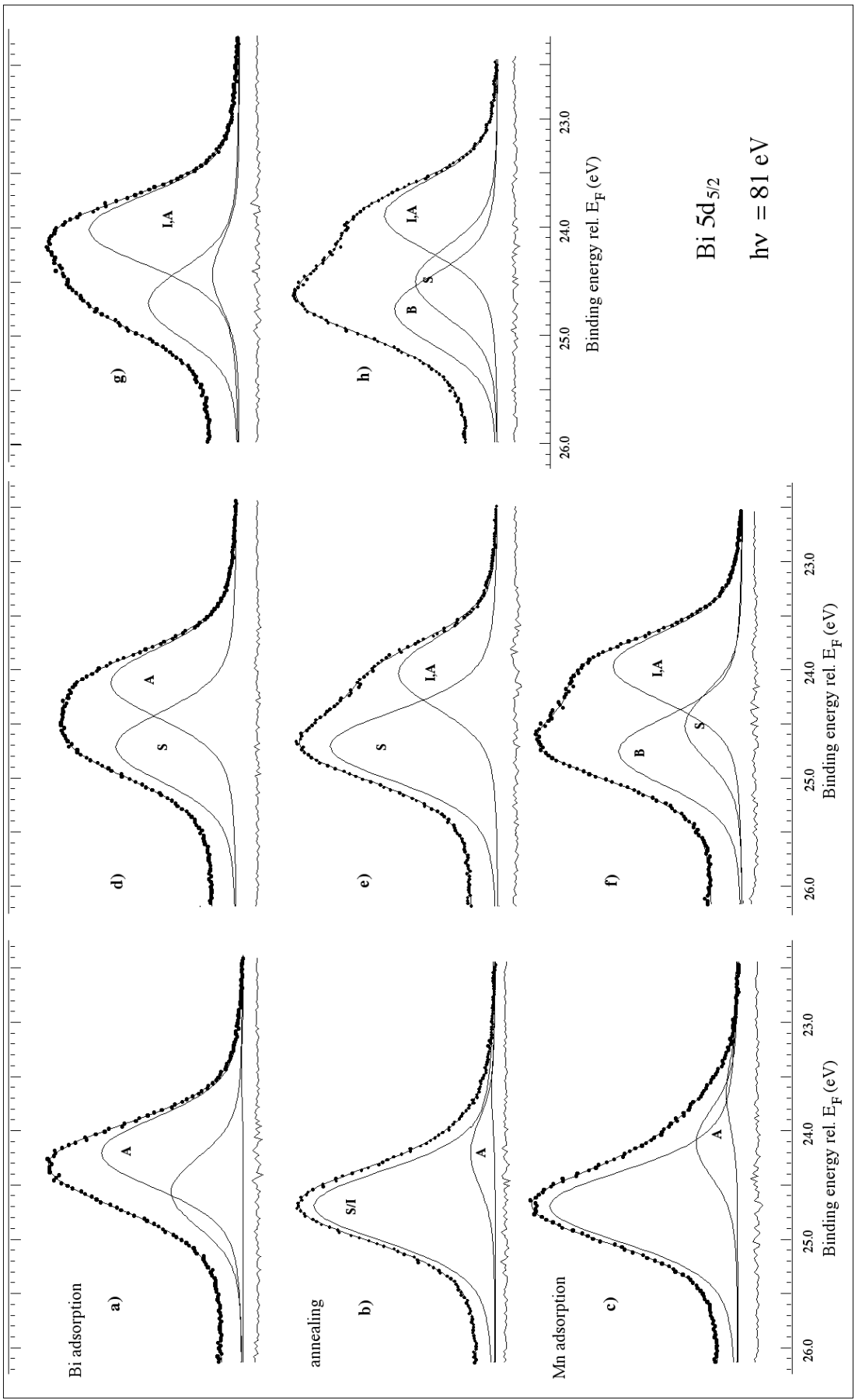

Figure 4

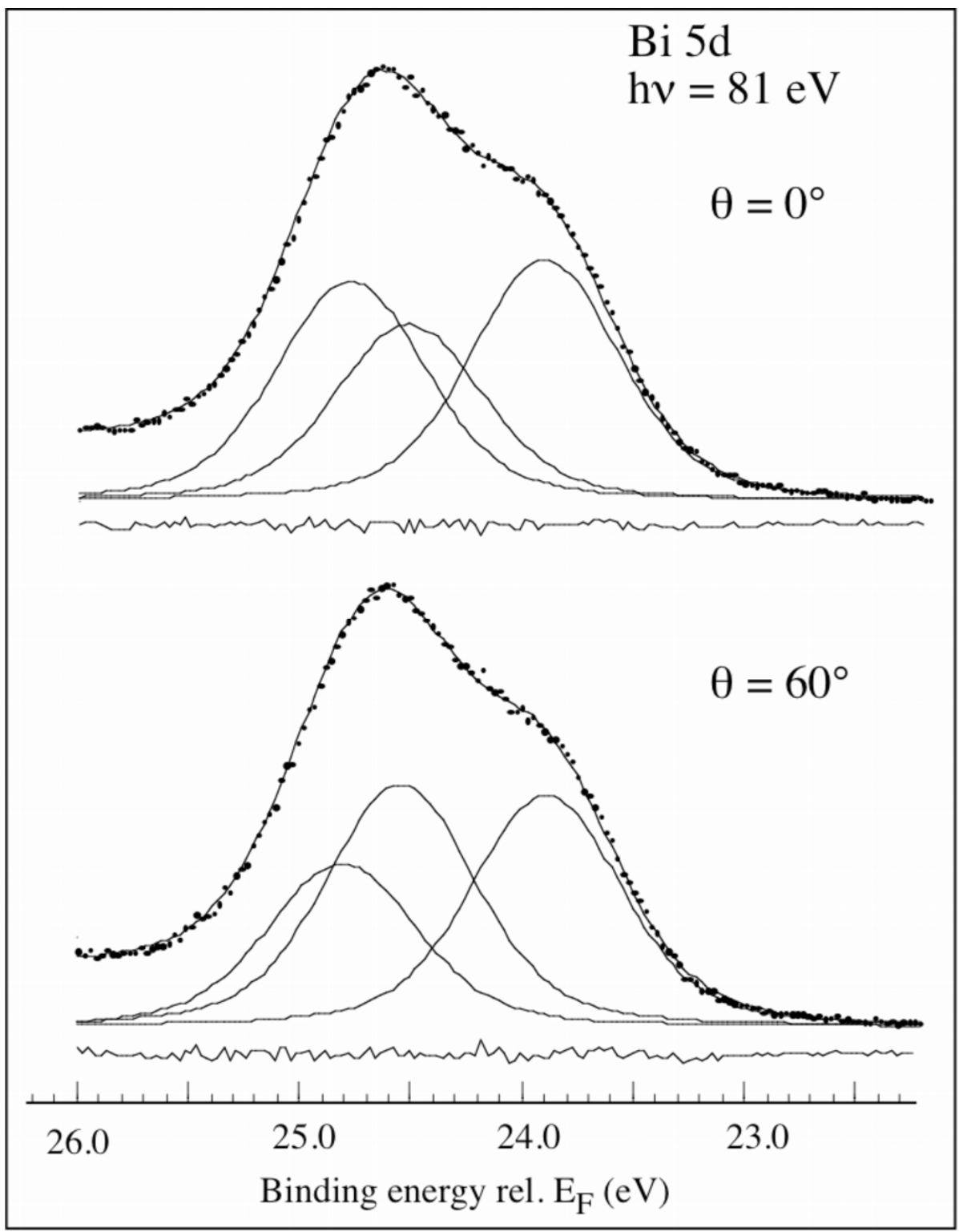

**Figure 5**